\pdfoutput=1 

\documentclass[
    ,final            
    ]
  {aipproc}

\layoutstyle{8x11double}

\usepackage{amsmath,amssymb,amsfonts}
\usepackage[utf8]{inputenc}

\newcommand{\un}[1]{\ensuremath{\ \mathrm{#1}}}

\newcommand{\rsun}{\ensuremath{\ R_\odot}} 
\newcommand{\rot}[1]{\ensuremath{\nabla\times {#1}}}
\newcommand{\diver}[1]{\ensuremath{\nabla\cdot {#1}}}

\begin{document}

\title{Coronal Inflows and Giant Polar Plumes}

\classification{96.60.-j,96.60.P-,96.60.pc,96.60.Vg}
\keywords      {Sun, MHD, Polar plumes, Coronal jets, Solar wind}

\author{R. Pinto}{
  address={Observatoire de Paris, LUTH, CNRS, 92195 Meudon, France}
  , altaddress={Institut d'Astrophysique Spatiale, Université Paris
    XI, Orsay, France},
  email={rui.pinto@obspm.fr}
}

\author{R. Grappin}{
  address={Observatoire de Paris, LUTH, CNRS, 92195 Meudon, France}
}

\author{J. Léorat}{
  address={Observatoire de Paris, LUTH, CNRS, 92195 Meudon, France}
}

\begin{abstract}
  We present the first results of simulations of giant polar plumes
  and coronal flows.
  We use a 2.5D axisymmetric MHD numerical model of an isothermal
  corona and slow solar wind. A plume is generated just above a
  small magnetic bipole embedded in an unipolar flux region which is
  perturbed by Alfvén waves injected from the coronal base. 
  The boundary conditions are transparent.
  The results are compared to those obtained previously with a 1D wind
  model in which plumes are generated as a 
  consequence of variations of the heating and flux-tube expansion
  parameters. 
\end{abstract}

\maketitle


\section{Introduction}

Polar plumes are jet-like overdense features seen over the
polar regions, mostly during the low solar activity phases. These
features extend radially, in the direction of the magnetic field in
the coronal holes. Plumes are found to
overlie EUV bright points, and these small bipoles appear to be
undergoing interchange reconnection with the unipolar flux
concentrations inside coronal holes. Outward propagating slow Alfvén
waves with periods between $10$ and $15\un{min}$ and amplitudes of about
$10\%-20\%$ of the plumes' intensity can be observed in
the first $\sim 0.2\rsun$ \citep{deforest_observation_1998}. 
\citep{ofman_slow_1999,ofman_dissipation_2000} studied the
dissipation of such waves and its contribution to the coronal heating
and solar wind acceleration.
We focus here in the study of the formation and decay of coronal plumes 
(and associated in/outflows), and report some preliminary results of
using an MHD numerical model of an isothermal axisymmetrical solar
corona and wind. 
Alfvén waves are meant to perturb a small magnetic
bipole initially in steady-state equilibrium  at the bottom of a
unipolar coronal hole. The waves generate unsteady pressure 
patterns which lead to current density accumulation and reconnection.
The results are compared to those 
found by \citep{pinto_time-dependent_2009} for a radially oriented
magnetic flux tube 
which includes the dense chromospheric layers and the solar wind.
In this later case, the wind is perturbed by rapid variations of the
heating rate in the chromosphere and low corona, producing transient
in and outflows before the system reaches a new steady-state equilibrium.
Both cases show the formation of an overdense stream which
progresses outwards in the form of a single wavefront with a phase
velocity much higher than the bulk stream velocity (of the order of
the surrounding wind speed).
The decay of such features is accompanied by transient inflows visible
mostly in the lower corona. 

\smallskip
\textit{Photospheric forcing.}
Slow photospheric movements (null frequency) are expected to shear
the coronal magnetic arcades by twisting their footpoints. This
``twisting'' may be transmitted upwards and so contribute to the 
magnetic tension/pressure accumulation, for as long
as the magnetic diffusion remains negligible. Finite frequency
movements, on the other hand, are more prone to 
inject waves (\emph{e.g} sound and Alfvén waves) into the very same
magnetic structures.
The standard numerical approach to the modelling of the effects of
such movements is that of the
rigid \emph{line-tied} forcing of the magnetic loops by their
footpoints, which are attached to the photosphere (both for the null
and the finite frequency cases). An important consequence of this
approximation is that total reflection occurs
at the foot-points of the loops, which then act like resonant
cavities; finite-frequency footpoint motions induce long-lasting and large
amplitude oscillations \citep{berghmans_coronal_1995} and null
frequency motions shear the loop by an arbitrarily high amount.
The \emph{line-tying} approximation is, nevertheless, one
among several possible approximations, and an extreme one in some
senses. \citep{grappin_mhd_2008} have shown that, at least in the case
of a single coronal loop perturbed by  ``slow'' photospheric
movements, the line-tying approximation may be inappropriate to
characterise the coronal loop's dynamical response.
In what follows, we will adopt an alternative approximation which
diametrically opposes the \emph{line-tying} one: we will consider an
isothermal MHD corona with a \emph{transparent} lower boundary (we'll
name it the \emph{fully transparent} case hereafter, for
simplicity).
We wish to stress at this point that both of these are extreme
approximations, the physical reality laying somewhere in between. The
\emph{line-tied} case is prone to severely overestimate the energy
input from the photosphere into the corona within magnetically connected
regions, and the \emph{fully
  transparent} case, on the other hand, underestimates it as the
absence of the denser chromospheric (and photospheric) layers reduces
the effective reflection ratios for waves crossing the lower
regions. \citep{gruszecki_attenuation_2007,gruszecki_influence_2008}
studied the wave energy leakage through these layers which, despite
being non-negligible, may be smaller than other forms of energy dissipation.
The rationale behind the present model (which uses the \emph{fully transparent}
approximation) is to consider a ``pessimistic''
configuration in what concerns the destabilisation of coronal
structures and search for mechanisms which nonetheless are able to produce
coronal events such as polar plumes, jets and inflows. More realistic
(less extreme) situations will be considered in future work.

\smallskip
\textit{Heating.}
The ohmic dissipation and consequent heating will not be taken into
account in our isothermal model, but the generation and accumulation of
current density $J$ can still be quantified. The analysis of
the evolution of $J$ in the domain may give clues to bridge and/or
compare our results to others from non isothermal models.


\section{Giant Polar plume}

\textit{Numerical model.}
\begin{figure}
  \includegraphics[width=.9\linewidth]{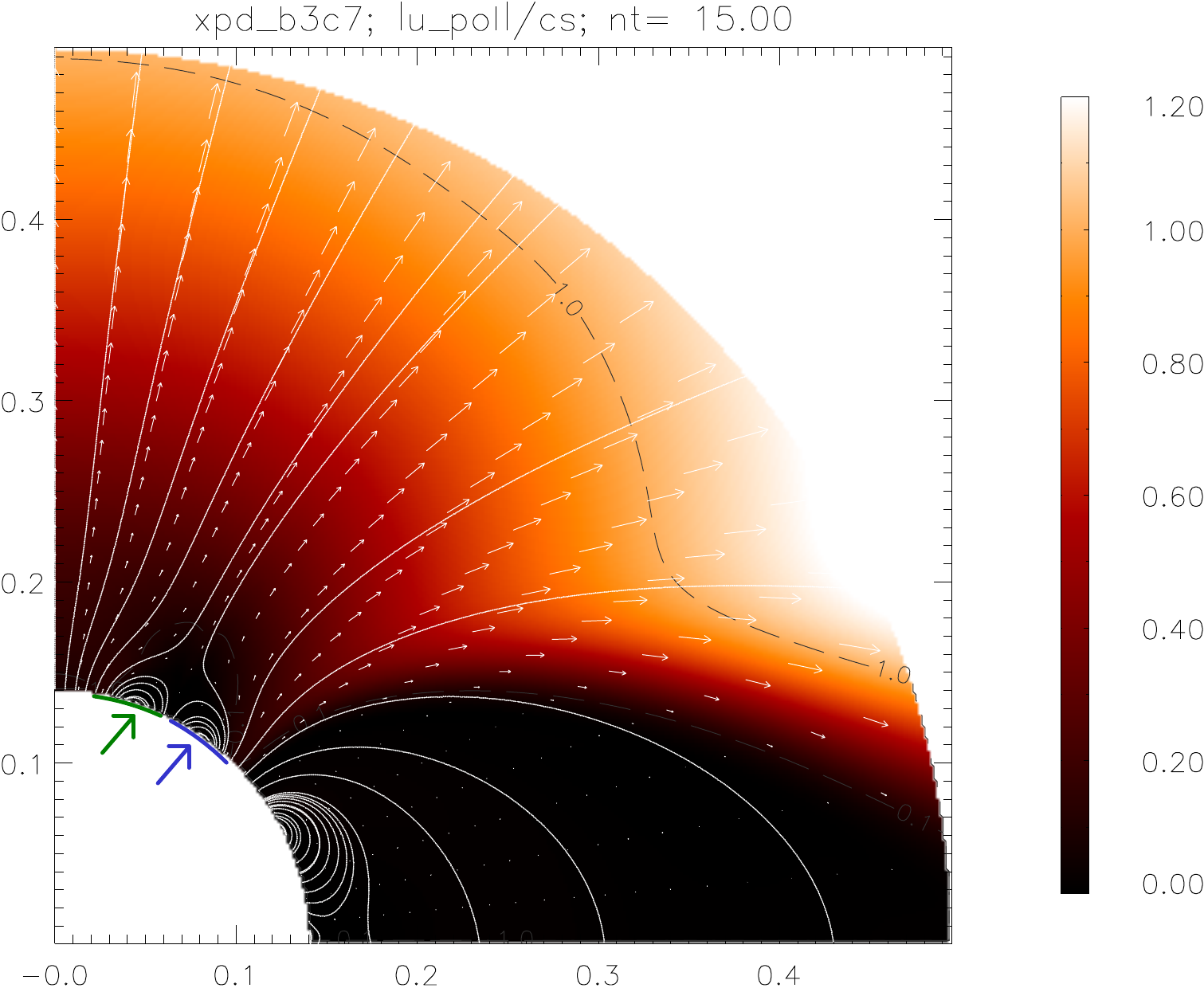}
  \caption{Initial conditions. Only the northern hemisphere is
    shown, and the radial sub-domain $1 \leq r/R_\odot \leq
    3.5$. Color-scale represents the poloidal Mach number, 
    the dashed contours levels $\left|u_{pol}\right|/c_s=0.1, 1.$ 
    (with $c_S\approx 150\un{km/s}$) and the white
    lines are magnetic field-lines. The large arrows show
    where Alfvén waves will be injected.} 
  \label{fig:ciplume}
\end{figure}
\begin{figure}
  \centering
  \includegraphics[width=0.8\textwidth]{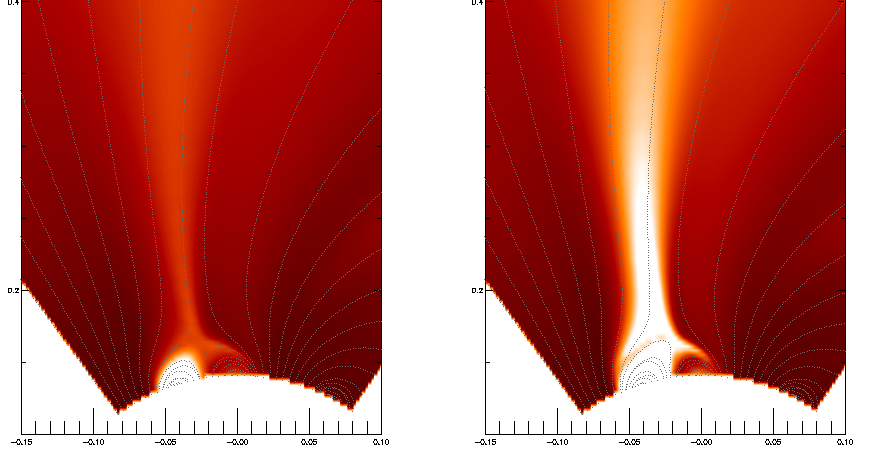}
  \includegraphics[width=0.2\textwidth]{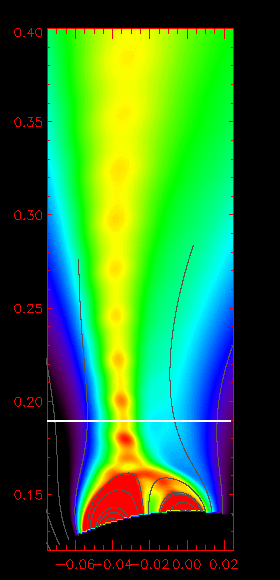}
  \caption{Formation of a jet above the bipolar structure in
    Fig. \ref{fig:ciplume} (turned $30^\circ$ CCW). The colour-scale
    represents $\rho$ corrected to compensate for the stratification
    in the first $r\sim3\rsun$ of the domain. The time intervals
    after the beginning of the wave injection are, from left to right,
    $2.5,\ 4.2,\ 5.9$ and $7.5\un{h}$. The amplitude of the injected
    wave is $u_\phi^0\approx 49\un{km/s}$. The rightmost smaller
    image shows the instant $t=7.5\un{h}$ with tighter colour
    ranges.}
  \label{fig:colunas}
\end{figure}
We use an 2.5 D axisymmetric model of the solar corona obeying
the following one-fluid isothermal MHD equations
\begin{eqnarray*}
  \label{eq:mhd}
  \partial_t \rho & + & \diver{\rho\mathbf{u}} =0\ ;\ \  P = \frac{2}{m_H}\rho k_BT \\
  \partial_t \mathbf{u} & + & \left(\mathbf{u}\cdot\nabla\right)
  \mathbf{u} = -\frac{\nabla P}{\rho} +
  \frac{\mathbf{J}\times\mathbf{B}}{\mu_0\rho} -
  \mathbf{g} + \nu\nabla^2\mathbf{u} \\ 
  \partial_t\mathbf{B} &=& \rot{\left(\mathbf{u}\times\mathbf{B}\right)} +
  \eta\nabla^2\mathbf{B}
\end{eqnarray*}
The magnetic field $\mathbf{B}$ decomposes into a time-independent
\emph{external} component $\mathbf{B^0}$ and an induced one
$\mathbf{b}$. The solar wind develops into a stable transsonic solution
in the open field regions. Both the upper and the lower boundaries (respectively at
$r=14$ and $1.01\rsun$) are transparent. Alfvén waves are injected by
perturbing the corresponding characteristic there. 
The diffusive terms are adapted so that grid scale ($\Delta l$)
fluctuations are correctly damped. The kinematic viscosity is defined as
$\nu=\nu_0\left(\Delta l/\Delta l_0 \right)^2$, typically with
$\nu_0=2\times 10^{14}\un{cm^2\cdot s^{-1}}$ and
$0.01\lesssim\left(\Delta l/\Delta l_0 \right)^2\lesssim10$. 
The resistive term is replaced by an implicit filter which 
dissipates mostly at the grid scale and minimises the dissipation of
large scales fluctuations. Note that actual kinetic dissipation
should happen at scales much smaller than the grid size, so this
approximation isn't less realistic than a laplacian term.
The principals of
the numerical model are thoroughly discussed in
\citep{grappin_alfvn_2000}.
The focus here is on circumventing the limitations found in
\citep{grappin_mhd_2008} by means of a convenient choice of magnetic
configuration (by setting $\mathbf{B^0}$) and Alfvén wave
injection domain. 
In particular, we chose to perturb regions of the surface not
directly connected by magnetic field lines, and we do so by
starting from a magnetic configuration with a null point
and separate connectivity domains. This way, we can expect wave mode
coupling to happen due to the highly non-uniform phase velocity
distributions, the existence of $C_s=C_a$ interfaces and very large
Alfvén crossing times near the null 
point \citep{mclaughlin_mhd_2006,landi_alfvn_2005}.

\smallskip
\textit{Initial conditions.}
A close-up of the northern polar region of the numerical
domain is displayed in Fig. \ref{fig:ciplume}. This particular
configuration is comparable to the pseudo-streamer
structures such as those in \citep{wang_solar_2007}.
Monochromatic Alfvén waves
will be injected in the zones indicated by the large
arrows in Fig. \ref{fig:ciplume}. The wave period is in the range
$T_0=6-35\un{min}$, the amplitude $u_\phi^0\approx 49\un{km/s}$ at
the bottom of the corona and each side will be
perturbed with a half period phase difference (such that
pseudo-streamer structure oscillates as a whole).
The plasma's $\beta$ is lower than $1$ all along the lower boundary,
and greater than $1$ only over and around the null point.


\smallskip
\textit{Formation of a jet.}
\begin{figure}
  \centering
  \includegraphics[width=\linewidth]{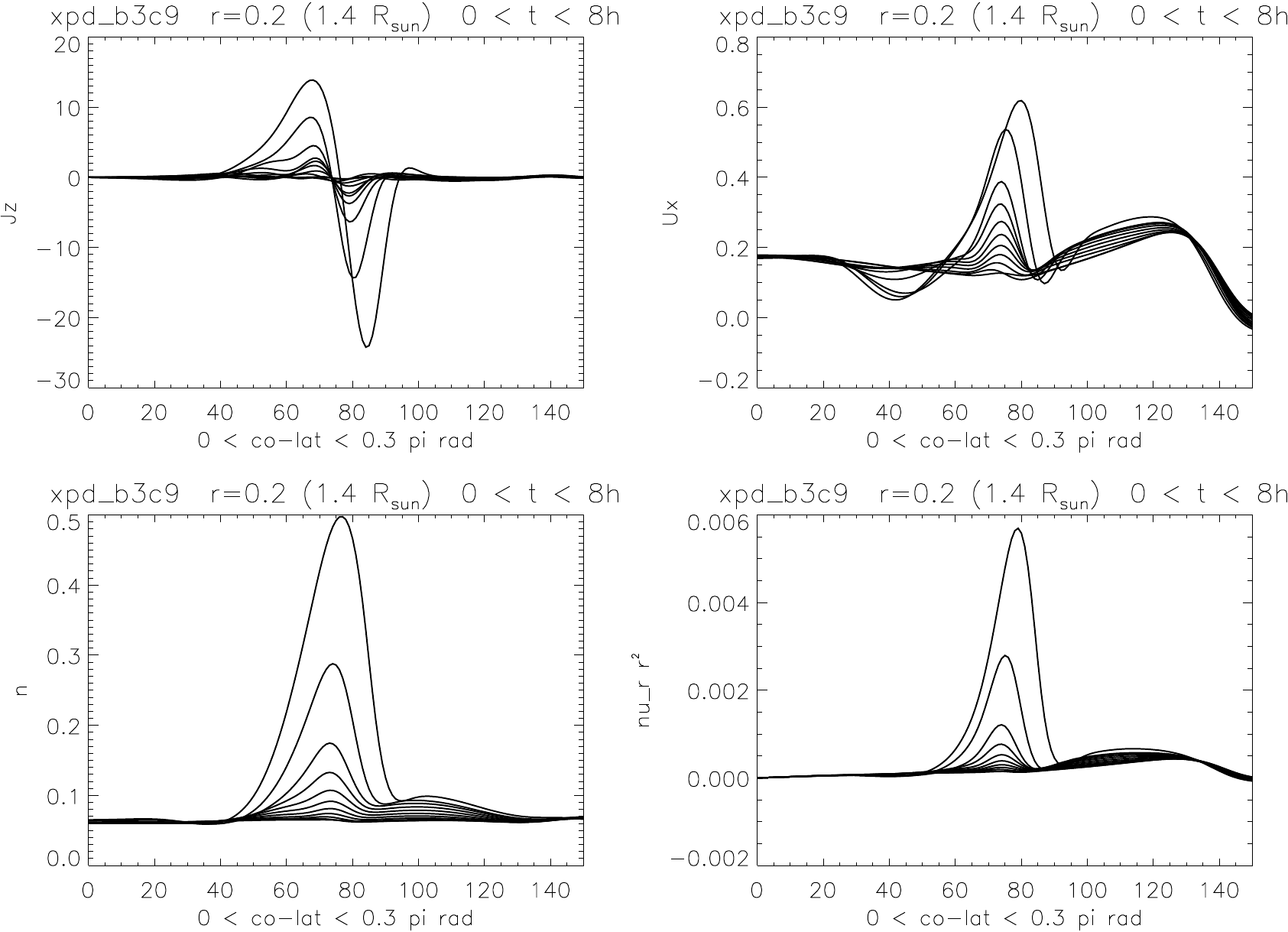}
  \caption{Profiles of $J_\phi,\ U_r,\ \rho,\ \rho U_r r^2$
    (top left, top right, bottom left and bottom right,
    respectively) over the white line in
    Fig. \ref{fig:colunas}. Different curves correspond to
    different instants with $\delta t \sim 50\un{min}$.}
  \label{fig:cortes}
\end{figure}
The Alfvénic wave fronts produce a systematic (yet fluctuating)
magnetic pressure gradient $\nabla b_\phi^2/2$ which push up the
denser layers into the $\beta>1$ region, force reconnection and plasma
diffusion up into the open field region just above the null point. An
over-dense plasma column forms continuously above the null point as
the diffused plasma joins the solar wind.
The jet presents a series of blobs, which correspond to density
enhancements of amplitude $\frac{\delta\rho}{\rho}<0.1$ which
propagate upwards with a phase velocity equal to $u_{//} + c_s$ as
slow mode wavefronts.
(Fig. \ref{fig:colunas}). The outermost edge of the jet progresses
also with a velocity equal to $u_{//} + c_s$, above the
jet's bulk velocity $u_{//}$. The plasma which does not join the jet
falls back along both sides of the pseudo-streamer, producing two
continuous inflows.
Fig. \ref{fig:cortes} shows the evolution of the current density,
radial velocity, density and mass flux over a cut represented as a white line
in Fig. \ref{fig:colunas}.
A current sheet forms near the null point and along the magnetic
separatrices as the plasma is compressed upwards. Current density peaks
around the null point. The density contrast between the over-dense
column and the background grows up to $\sim 10$ at the base of the jet
but fades away at higher heights ($\sim 3$ at $7\rsun$ and $\sim 1$ at
$10\rsun$). The plasma's velocity becomes higher than the surrounding
solar wind within the jet, and so does the mass flux.

\smallskip
\textit{Growth rate vs. frequency.}
\begin{figure}
  \centering
  \includegraphics[width=0.49\linewidth]{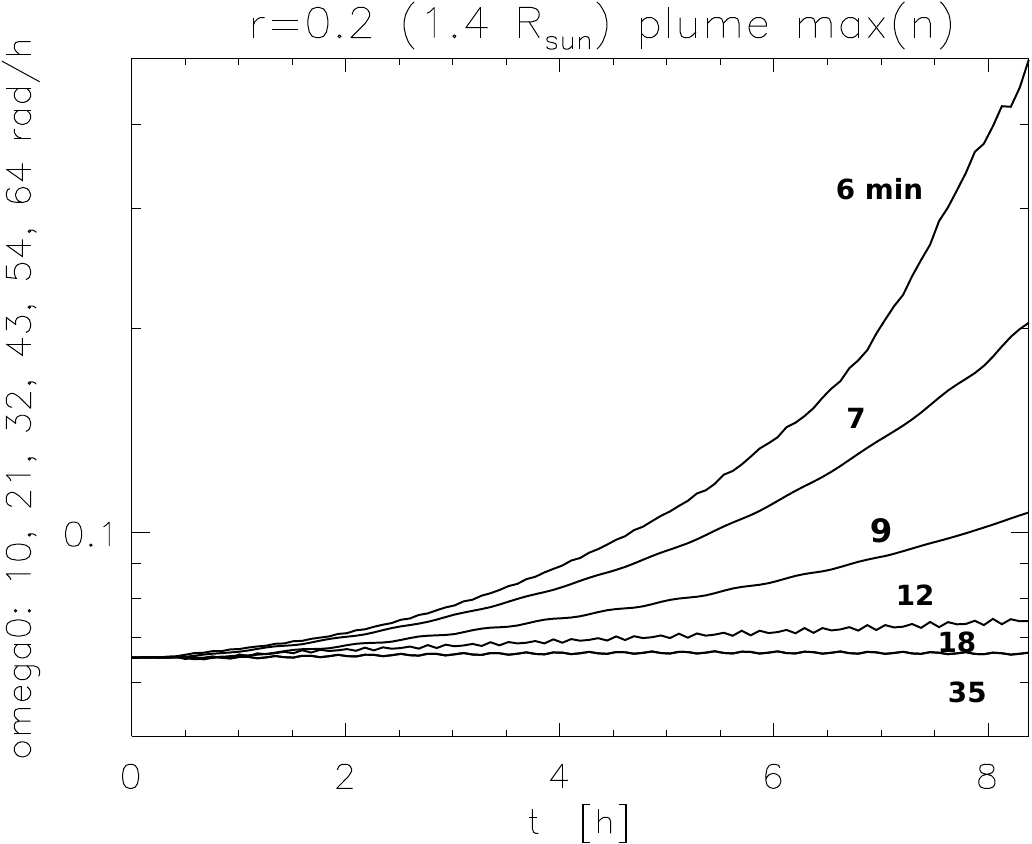}
  \includegraphics[width=0.49\linewidth]{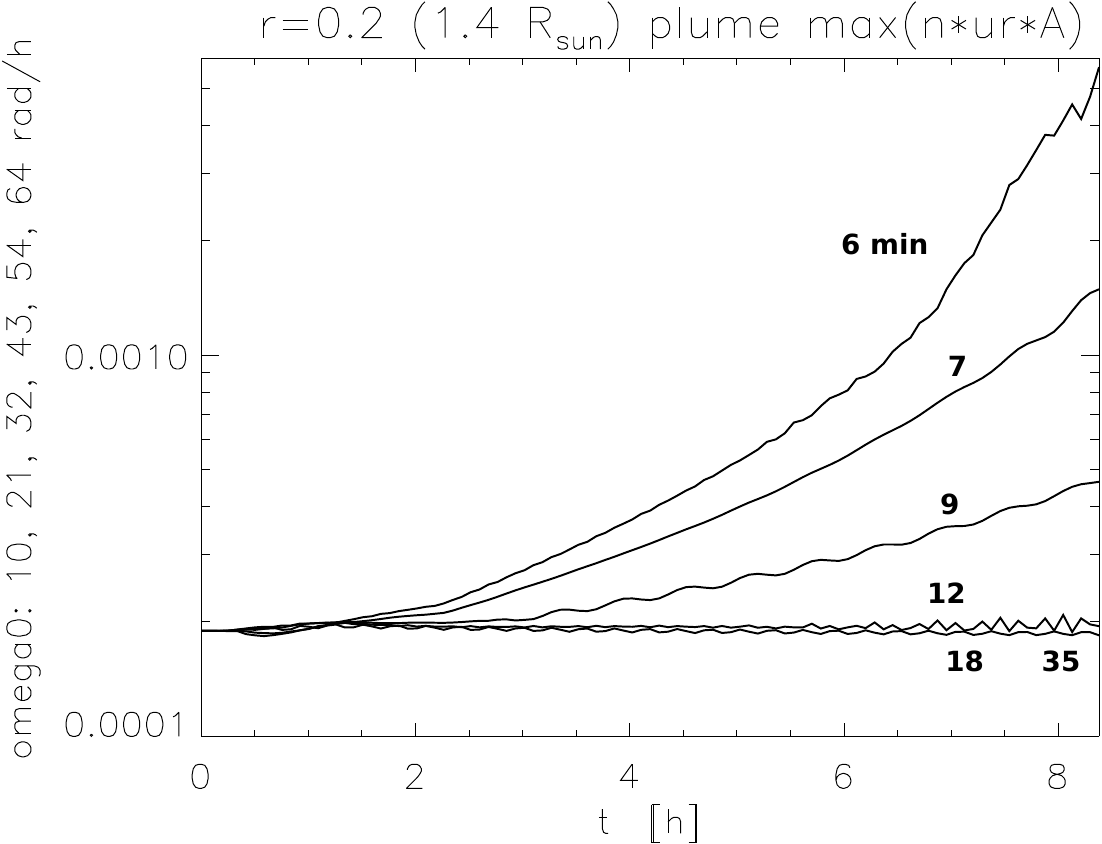}
  \caption{Density $n$ (left) and mass flux $n u_r A$ (right) as a
    function of time at the bottom of the jet ($r=0.2\sim
    1.4\un{\rsun}$, the same height as for the latitudinal cuts in
    Fig. \ref{fig:cortes}). Each curve corresponds to a given
    frequency $\omega_0$ (correspondent wave periods:
    $T_0=35,18,12,9,7,6\un{min}$) for the injected Alfvén waves. The
    higher the frequency, the higher the growth rate. The ordinates
    are in log-scale.}
  \label{fig:freq}
\end{figure}
The Alfvén wave driven overdense jet may either saturate and reach a
quasi-steady state or grow continuously for long periods of time,
depending on the frequency $\omega_0$ of the injected waves.
Fig. \ref{fig:freq} shows the growth of the mass flux and density
within the jet as a function of time for several wave frequencies
$\omega_0$. There is a smooth transition between a stable and
an unstable regime at about $T_0=\frac{2\pi}{\omega_0}=12\un{min}$; 
lower frequencies lead to lower (and eventually null)
growth rates while higher frequencies lead to ever higher growth
rates. This behaviour does not seem to break with time, at least for as
long as the run lasts ($\Delta t\sim 8.5\un{h}$).

\section{Discussion and conclusions}

\begin{figure}
  \centering
  \includegraphics[width=.9\linewidth]{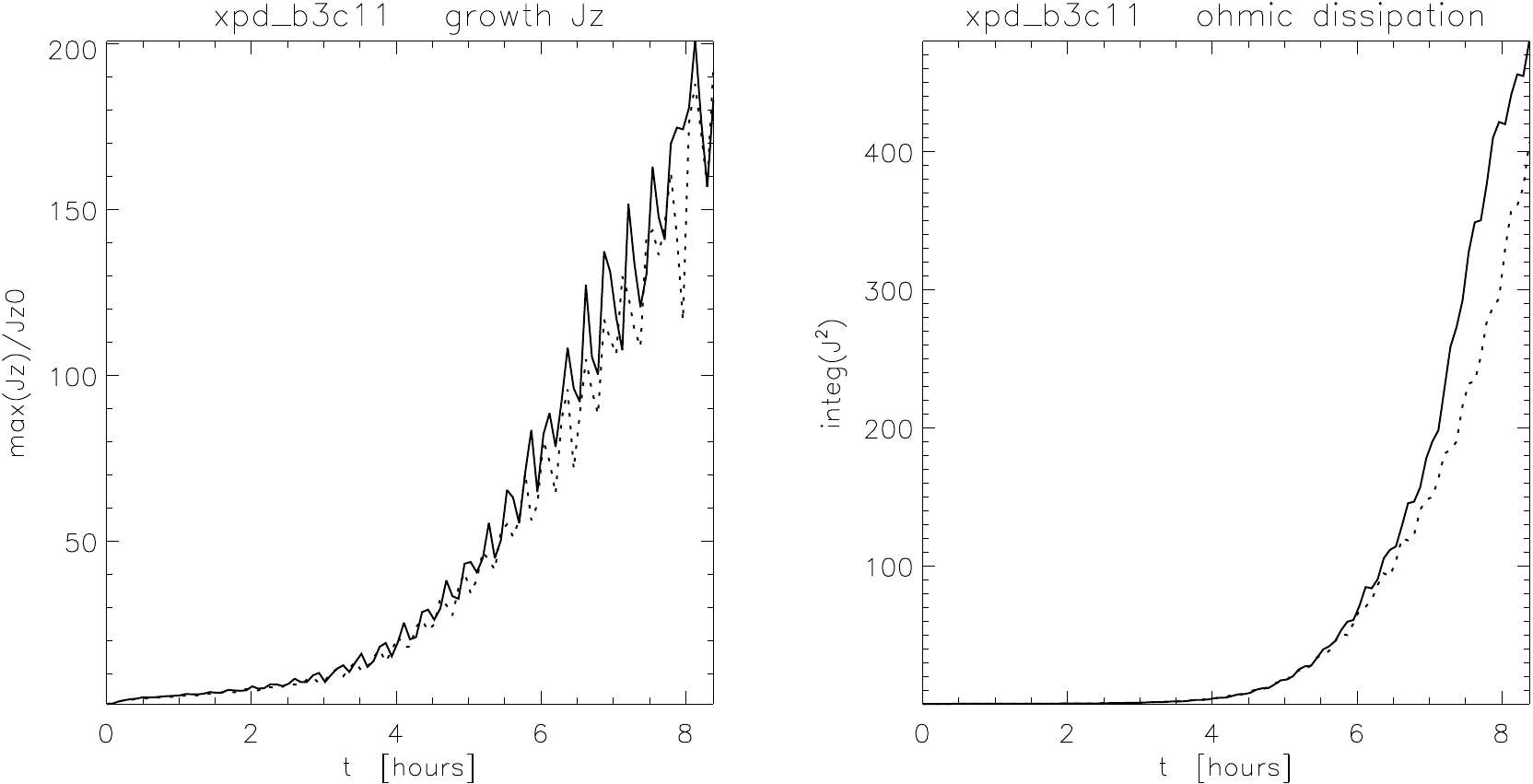}
  \caption{Left: peak current density $\max\left(J_\phi\right)/J_\phi^0$.
    Right: $\int_VJ^2_\phi dV$ in a volume $V$ which encloses the
    perturbed region (magnetic bipole plus the lower part of the
    jet). The dotted line corresponds to a kinematic viscosity $2\nu_0$.}
  \label{fig:ohmic}
\end{figure}

We studied the long-term effects of finite frequency oscillations
of the coronal footpoints of a pseudo-streamer placed inside a polar
coronal hole using an MHD numerical model of an isothermal corona. We
use transparent boundaries and show that perturbing such
magnetic structures at their footpoints with Alfvén
waves can lead to impressive large-scale phenomena.
Transverse twisting movements at the base of the bipolar structure
(Fig. \ref{fig:ciplume}) lead to a
long lasting destabilisation of the system and the formation of a
dense jet (Fig. \ref{fig:colunas}) which may be interpreted as a polar
plume. These movements are Alfvén waves injected at the base of the
 bipole, its two 
groups of magnetic arcades being perturbed in phase opposition one
relative to the other. The effective twisting angle of the whole structure
remains small (a few degrees) and oscillates with finite
frequency, unlike in line-tied photospheric shearing models
(e.g \citep{pariat_model_2009}).
A complex pattern of Alfvén wavefronts propagating within the lower
magnetic arcades
non-linearly produces density enhancements which are pushed upwards, a
part of which then diffuses across the magnetic field up into the open
field region. The diffused plasma then joins the overlying polar
plume. This diffusion shows a temporal modulation with a period
equal to that of the injected waves (a consequence of exciting
both magnetic arcades with exactly a half period phase difference;
choosing different phase relations leads to different frequencies). As
a result, the plume displays 
a series of blobs which propagate along its axis as slow mode
wavefronts. This result is consistent with the slow waves observed by
\citep{deforest_observation_1998}.
Current accumulates around the null point and along the magnetic
separatrices. Fig. \ref{fig:ohmic} highlights the growth of current
density $J_\phi$ as a function of time and provides a mesure of the
production of magnetic small scales. The ohmic heating could be
defined as $Q_{ohm}\left(t\right) =\eta \int_VJ^2 dV$, where $V$
encloses the perturbed region \citep{ofman_solar_1997}.
Note that the actual values of $Q_{ohm}$ depend on the resistive
settings and therefore on the local grid-scale, such that
$Q_{ohm}\propto \Delta l^{-1}$. Varying the kinematic viscosity, on the other
hand, as little effect on the result. Future work will bring
deeper insight over the dissipative phenomena.
The ohmic heating cannot be taken into account here (the
model is isothermal), but we assume that the simulations of
\citep{pinto_time-dependent_2009}, who use a 1D
non-isothermal model to study the dynamical effects of
varying the heating rate and flux-tube expansion in the lower
atmosphere, can represent
an open flux tube passing close to the magnetic bipole in
Fig. \ref{fig:ciplume}. The heating rate variations at
low coronal heights are a proxy to the current accumulation and
ohmic dissipation. Plume and interplume states
are also obtained this way, with similar predictions for formation
and decay time-scales and density contrast. On the other hand, the
velocity of the wind for the plume solution they find is smaller than
that for the interplume solution. To clear this point, we are working
on extending the thermal treatment to the 2D
case to have a unified (and self-consistent) view of these
aspects.


\begin{theacknowledgments}
  We thank the referee for useful suggestions. The numerical
  simulations were carried out using the IDRIS
  computing facilities.
\end{theacknowledgments}



\bibliographystyle{aipproc} 

\bibliography{mnemonic,refs}

\IfFileExists{\jobname.bbl}{} {\typeout{}
  \typeout{******************************************} \typeout{**
    Please run "bibtex \jobname" to optain} \typeout{** the
    bibliography and then re-run LaTeX} \typeout{** twice to fix the
    references!}  \typeout{******************************************}
  \typeout{} }

\end{document}